\documentclass[]{aa}
\usepackage{graphicx}
\usepackage{txfonts}
\usepackage{natbib}
\usepackage{ifthen}
\newcommand{\highlight}[1]{{#1}}

\newcommand{\ProgramOne}{190.C-963}
\newcommand{\ProgramTwo}{087.C-0709}

\title{A study of dust properties in the inner sub-au region\\ of the Herbig Ae star HD~169142
  with VLTI/PIONIER
    \thanks{
        Based on observations collected at the European Organisation for
        Astronomical Research in the Southern Hemisphere under ESO programs
        {\ProgramOne} and \ProgramTwo.
    }
}

\titlerunning{Dust properties in HD~169142}

\author{L. Chen\inst{1}
  \and \'A. K\'osp\'al\inst{1,2}
  \and P. \'Abrah\'am\inst{1}
  \and A. Kreplin\inst{3}
  \and A. Matter\inst{4}
  \and G. Weigelt\inst{5}
}

\institute{
Konkoly Observatoy, Research Centre for Astronomy and Earth Sciences,
  Hungarian Academy of Sciences,
\\ Konkoly Thege Mikl\'os \'ut 15-17, 1121 Budapest, Hungary
  \\email: lei.chen@csfk.mta.hu
\and Max-Planck-Institut f\"ur Astronomie, K\"onigstuhl 17, 69117 Heidelberg, Germany
\and University of Exeter, Department of Physics and Astronomy, Stocker Road, Exeter, Devon EX4 4QL, UK
\and Laboratoire Lagrange, Universit\'e C\'ote d'Azur, Observatoire de la C\'ote d'Azur, CNRS, Boulevard de l'Observatoire, CS 34229, 06304 Nice Cedex 4, France
\and Max-Planck-Institut f\"ur Radioastronomie, Auf dem H\"{u}gel 69, 53121 Bonn, Germany
}

\newcommand{\GM}{geometric model}
\newcommand{\RT}{radiative transfer}
\newcommand{\MCRT}{Monte-Carlo radiative transfer}

\newcommand{\SED}{spectral energy distribution}

\newcommand{\enDash}{{\textrm{--}}}
\newcommand{\ParameterRange}[2]{$[#1,~#2]$}
\newcommand{\IR}{infrared}
\newcommand{\NIR}{near-infrared}

\newcommand{\Kelvin}{\mathrm{K}}
\renewcommand{\sun}{\odot}
\newcommand{\degree}{\degr} % \degr is defined in aa.cls
\newcommand{\mas}{{\mathrm{mas}}}
\newcommand{\AU}{{\mathrm{au}}}
\newcommand{\pc}{{\mathrm{pc}}}
\newcommand{\mum}{{\mu\mathrm{m}}}

\newcommand{\meter}{{\mathrm{m}}}

\begin{document}
\newboolean{RefereeMode}
\setboolean{RefereeMode}{true}
\setboolean{RefereeMode}{false}

\newlength{\ColumnWidth}
\setlength{\ColumnWidth}{ 9.0cm}

\iftrue
\abstract
{%Context 
An essential step to understanding protoplanetary evolution
is the study of
disks that contain gaps or inner holes.
The pre-transitional disk around the Herbig star HD~169142
exhibits
multi-gap disk structure,
differentiated gas and dust distribution,
\highlight{planet candidates},
and {\NIR} fading in the past decades,
which make it a valuable target for a case study
of disk evolution.
}
{%Aims
Using {\NIR} interferometric observations with VLTI/PIONIER,
we aim to study the dust properties in the inner sub-$\AU$ region of the disk
 in the years 2011-2013,
when the object is already in its {\NIR} faint state.
}
{%Methods
  We first performed simple {\GM}ing to characterize the size and shape of the NIR-emitting region.
    We then performed {\MCRT} simulations on grids of models
    and
    compared the model predictions
    with the interferometric and photometric observations.
}
{%Results
We find that
the observations are consistent with optically thin gray dust
lying at $R_\mathrm{in} \sim 0.07~\AU$,
passively heated to $T\sim1500~\Kelvin$.
Models with sub-micron optically thin dust are excluded
because such dust will be heated to much higher temperatures at similar distance.
The observations can also be reproduced with
a model consisting of optically thick dust at $R_\mathrm{in} \sim 0.06~\AU$,
but this model is plausible only if refractory dust species enduring ${\sim}2400~\Kelvin$ exist in the inner disk.
}
{%Conclusions
}
\fi


\maketitle

\section{Introduction}
\iftrue

An essential step to understand{{ing}} protoplanetary evolution
is the study of
disks that contain gaps or inner holes.
These objects are likely to be in a late evolutionary stage
and are probably progenitors of debris disks.
The existence of gaps and holes may be inferred from the {\SED} (SED) in
infrared (IR).
Protoplanetary disks around Herbig Ae stars
are classified into two groups depending on the shape of their SEDs
\citep{2001A&A...365..476M}.
While a group II object has roughly a power-law SED from near-infrared (NIR) to FIR,
the SED of a group I object has an additional cold component at mid-infrared (MIR) or far-infrared (FIR) wavelengths.
A recent explanation
of the two types
is that
group I SEDs
are caused by disk gaps,
while group II objects typically have continuous disks
(\citealt{2013A&A...555A..64M}).
However,
\citet {2015A&A...581A.107M} suggested that group II objects
can have gaps that are narrow enough
not to cause appreciable signatures in the SEDs.
Gaps and inner holes have been directly detected
with spatially resolved observations
in a growing number of protoplanetary disks
\citep[e.g.,][]{2016PhRvL.117y1101I,2013Sci...340.1199V}.

The Herbig Ae star HD~169142 exhibits a typical group I SED,
with strong evidence for the presence of wide gap regions in its disk.
With direct MIR imaging,
\citet{2012ApJ...752..143H}
found an inner cavity
with a size of ${\sim}23~\AU$
in the MIR image of the disk.
However, the NIR excess indicates that hot dust
still exists
around the central star.
\citet{2012ApJ...752..143H}
therefore suggested that HD~169142 has a wide gap
separating a hot inner disk from a colder outer disk,
which would make this object
``an excellent candidate to look for newly formed planets in the protoplanetay disk".
Polarimetric imaging in the NIR of the object
\citep{2013ApJ...766L...2Q}
not only confirmed the cavity inside ${\sim}20~\AU$,
but also found a 
second gap
at $40\enDash70~\AU$,
just behind the inner rim of the outer disk.
Recent NIR polarized imaging
\citep{2015PASJ...67...83M,2017ApJ...838...20M}
 confirmed the double-gap disk structure.
These discoveries show HD~169142 to be among the first objects in which multiple gaps have been found.
A point-like IR source, possibly a brown dwarf or a forming planet,
was found at ${\sim}23~\AU$ from the central star,
that is, just at the inner rim of the outer disk
\citep{2014ApJ...792L..22B,2014ApJ...792L..23R}.
With radio observations at $7~\mathrm{mm}$, \citet{2014ApJ...791L..36O}
confirmed that the second gap is \highlight{likely to be caused by a dip in surface density profile},
and found a possible point source in this gap region.
Very recently,
imaging at $7~$mm and $9~\mathrm{mm}$
revealed another gap at ${\sim}85~\AU$
\citep{2017ApJ...838...97M}.
A decoupling of dust and gas components was found with the latest ALMA observation \citep{2017A&A...600A..72F}.
The dust continuum map exhibits similar gap regions as those in the NIR polarized light,
indicating that, instead of a shadowing effect,
the gap regions are indeed related to
a dip in surface density profile of dust.
However, the CO line intensity map indicates that gas components exist continuously in the gap regions.

\citet{2015ApJ...798...94W} found a decrease of 45\% in the NIR excess of HD~169142
over a timescale of not more than 10 yr,
from a pre-2000 high state to a post-2000 low state.%
They interpreted the variability as caused by
changes in the dust distribution
of an optically thin inner component,
while the optically thick core in their model remains constant
during the period of NIR variability.

The multi-gap disk structure,
differentiated gas and dust distribution,
\highlight{planet candidates},
and the variation in the inner disk
make HD~169142 a valuable target for a case study
of disk evolution.
However, 
spatially resolved observations of the inner disk,
where the most rapid evolution is happening,
are still rare.
Recently, {\IR} interferometric observations of HD~169142
with VLTI/PIONIER,
which for the first time resolved the sub-$\AU$ scale inner disk,
were reported in \citet{2017A&A...599A..85L},
as part of a large survey project.

In this paper we present both
refined {\GM}ing of HD~1619142 using a larger data set,
and our {\MCRT} modeling %of the data set
using RADMC3D%
\footnote{
http://www.ita.uni-heidelberg.de/~dullemond/software/radmc-3d/}.
Our study aims to constrain the dust properties of the inner disk.
Because all the NIR interferometric observations used in this work
were performed between the years 2011 and 2013,
our study constrains the dust properties in this period alone
when the object was already in its low state
\citep{2015ApJ...798...94W}.

In our study we adopt a distance of $117\pm4~\pc$,
derived from the recently measured parallax of $8.53\pm0.29~\mas$
for HD~169142 in Gaia~DR1
\citep{2016A&A...595A...2G}.
We note that a distance of $d=145~\pc$ was used in most of the previous studies
\citep[e.g.,][]{
2012ApJ...752..143H,%
2015ApJ...798...94W%
},
but most of our results are scalable and physically unaltered when adopting either distance value.

\fi

\section{Observations and data reduction}
\iftrue

\begin{figure}
  \includegraphics[width=9cm]{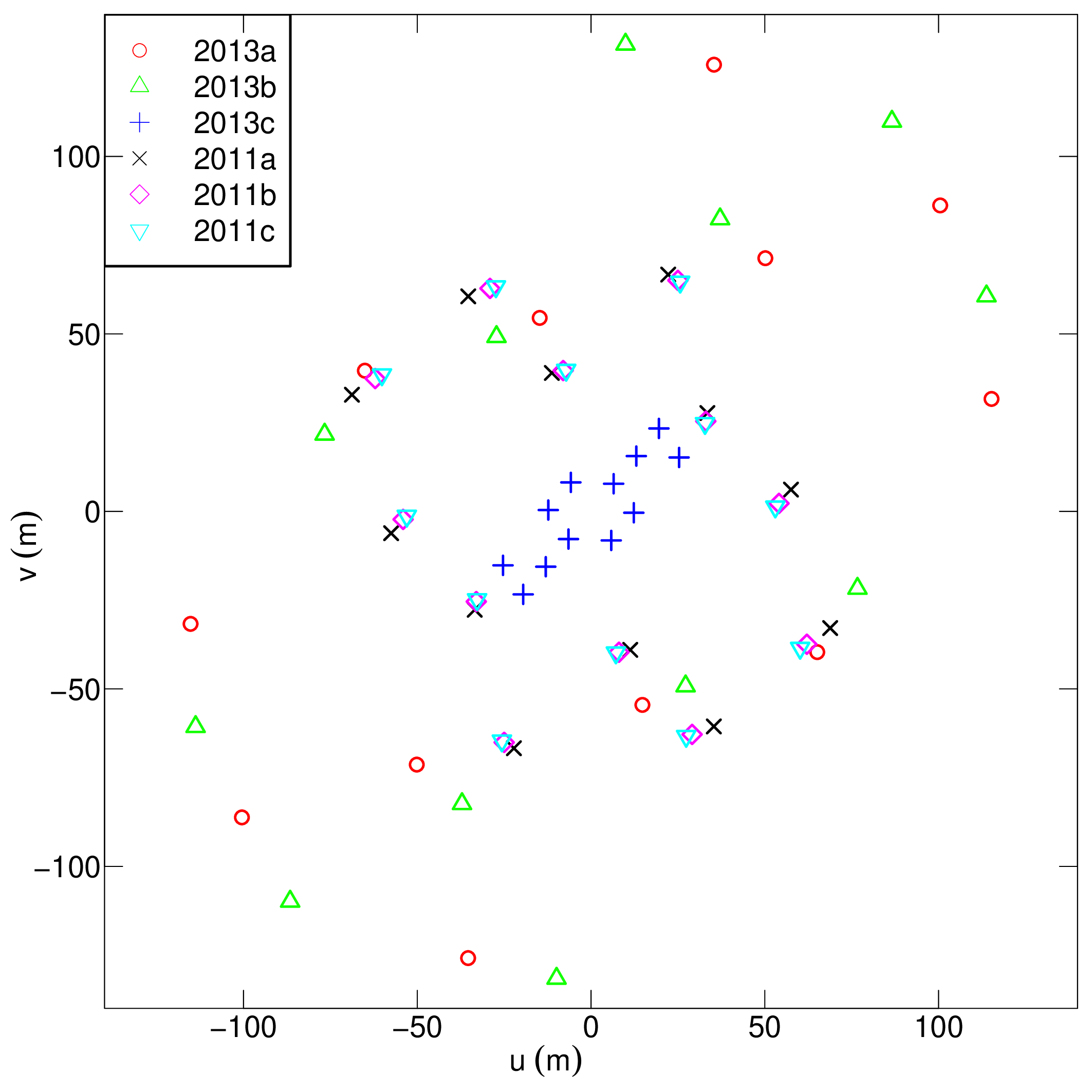}
  \caption{
    \highlight{$uv$-coverage of our interferometric observations.}
  }
  \label{fig:uv.coverage}
\end{figure}

\begin{figure}
  \includegraphics[width=9cm]{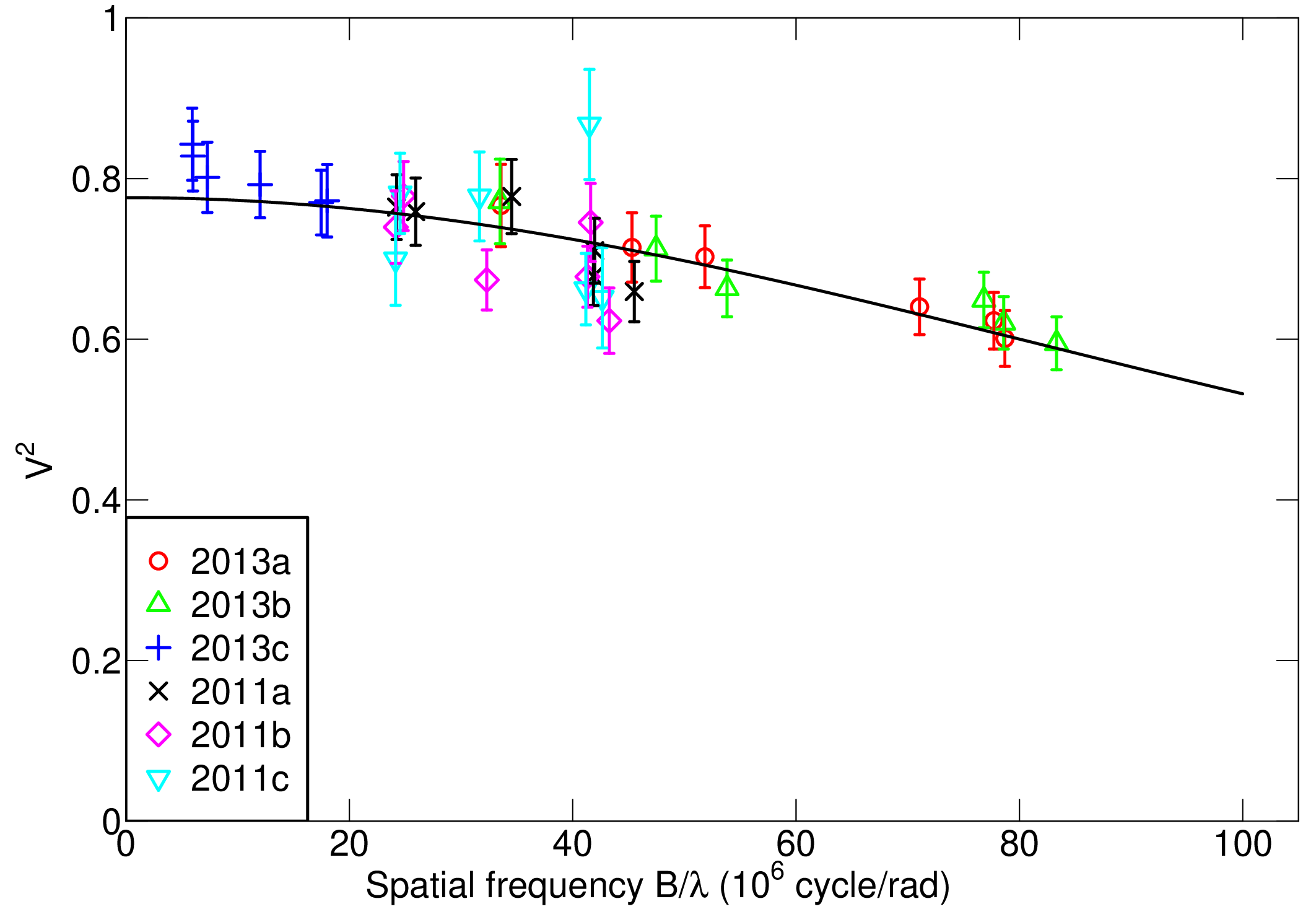}
  \caption{
    Visibilities from the PIONIER observations of HD~169142,
    and our star-ring-halo model fitting.
  }
  \label{fig:GM}
\end{figure}

\begin{figure}
  \includegraphics[width=9cm]{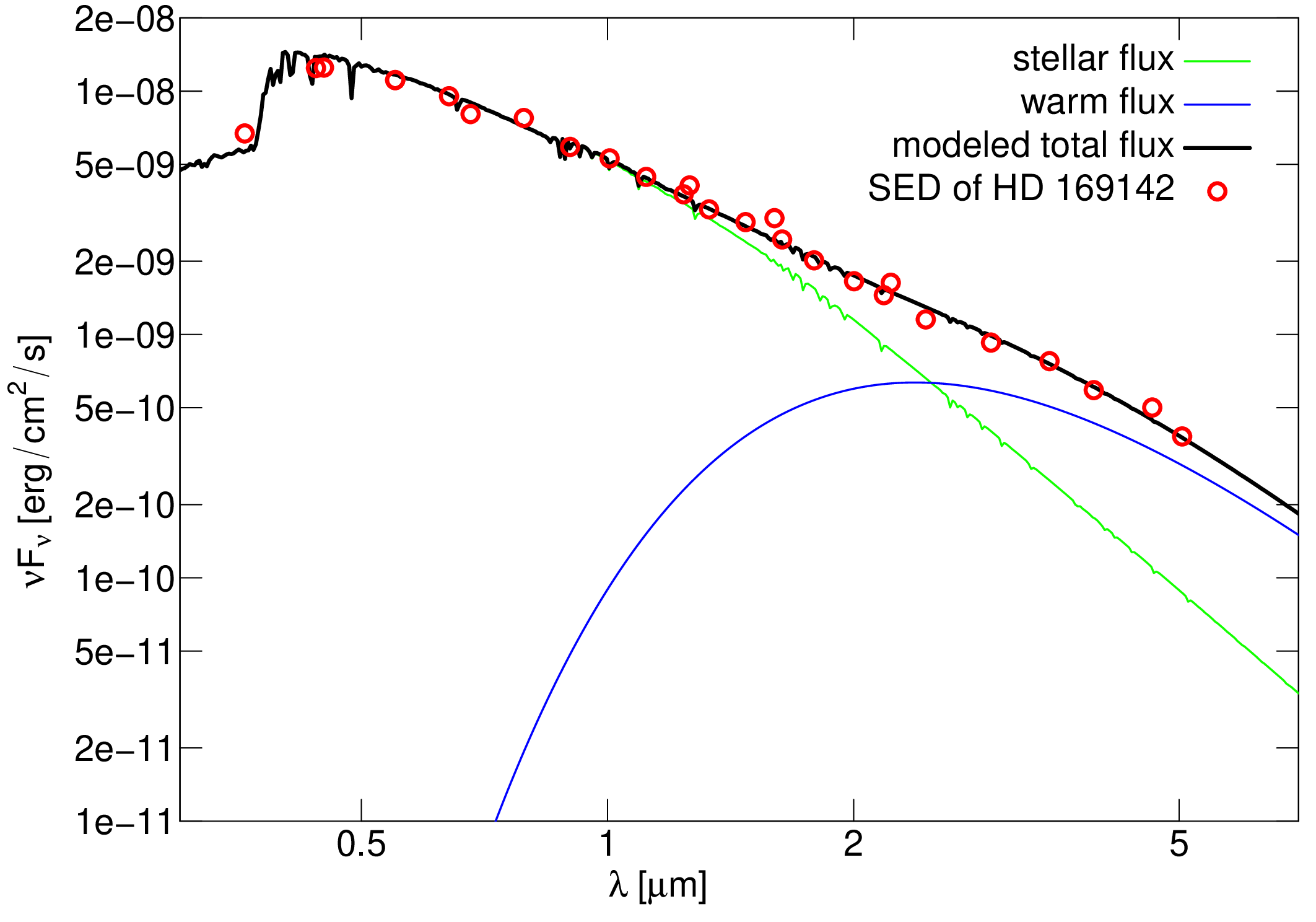}
  \caption{
    SED of HD~169142,
    and our Kurucz plus blackbody fitting.
  }
  \label{fig:SED}
\end{figure}

\begin{table*}
\caption{Observation log of our VLTI/PIONIER observations of HD~169142.
  For each observation, the information on the six baselines are shown.
  $B_p$ is the projected baseline length.
  PA is the position angle from north to east.
}
\label{table:observation.log}
\centering
\begin{tabular}{cccrrrrcc}
\hline
\hline
Data set  & Night  & Telescope configuration & $u$ & $v$ & $B_{\rm p}$ & PA
\\
 &             &      & (m)        & (m)          & (m)         & ($^\circ$)
\\
  \hline

2013a & 2013-06-04 & A1-G1-J3-K0 &  -14.8 &  54.5 &  56.5 &  -15.2\\
 &  &  &  -50.2 & -71.3 &  87.2 & -144.9\\
 &  &  &   65.1 & -39.6 &  76.2 &  121.3\\
 &  &  &  100.5 &  86.2 & 132.4 &   49.4\\
 &  &  & -115.2 & -31.7 & 119.5 & -105.4\\
 &  &  &   35.4 & 125.8 & 130.7 &   15.7\\
\hline

2013b & 2013-06-09 & A1-G1-J3-K0 &  -37.1 & -82.4 &  90.3 & -155.7\\
 &  &  & -113.8 & -60.7 & 128.9 & -118.1\\
 &  &  &    9.9 & 131.6 & 131.9 &    4.3\\
 &  &  &   76.7 & -21.7 &  79.7 &  105.8\\
 &  &  &   86.6 & 109.9 & 139.9 &   38.2\\
 &  &  &  -27.2 &  49.2 &  56.2 &  -28.9\\
\hline

2013c & 2013-07-03 & A1-B2-C1-D0 & -25.3 & -15.2 & 29.5 & -121.0\\
 &  &  &  12.3 &  -0.4 & 12.3 &   91.8\\
 &  &  &  -6.5 &  -7.8 & 10.1 & -140.2\\
 &  &  & -13.0 & -15.6 & 20.3 & -140.1\\
 &  &  &   5.8 &  -8.2 & 10.0 &  144.6\\
 &  &  & -19.5 & -23.4 & 30.5 & -140.1\\
\hline

2011a & 2011-06-05 & D0-G1-H0-I1 & 68.8 & -32.9 & 76.3 & 115.5\\
 &  &  & 22.2 &  66.7 & 70.3 &  18.4\\
 &  &  & 33.5 &  27.7 & 43.5 &  50.4\\
 &  &  & 57.5 &   6.1 & 57.9 &  83.9\\
 &  &  & 11.3 & -39.0 & 40.6 & 163.9\\
 &  &  & 35.3 & -60.6 & 70.1 & 149.7\\
\hline

2011b & 2011-06-05 & D0-G1-H0-I1 & 62.1 & -37.4 & 72.5 & 121.1\\
 &  &  & 25.0 &  65.1 & 69.7 &  21.0\\
 &  &  &  8.0 & -39.7 & 40.5 & 168.6\\
 &  &  & 29.1 & -62.8 & 69.2 & 155.2\\
 &  &  & 54.1 &   2.3 & 54.1 &  87.6\\
 &  &  & 33.0 &  25.4 & 41.7 &  52.4\\
\hline

2011c & 2011-06-05 & D0-G1-H0-I1 &  7.1 & -39.8 & 40.4 & 169.8\\
 &  &  & 25.6 &  64.6 & 69.5 &  21.6\\
 &  &  & 27.4 & -63.3 & 69.0 & 156.6\\
 &  &  & 60.2 & -38.5 & 71.4 & 122.6\\
 &  &  & 32.8 &  24.8 & 41.1 &  52.9\\
 &  &  & 53.0 &   1.3 & 53.0 &  88.6\\
\hline

  2011d & 2011-06-05 & D0-G1-H0-I1 & \multicolumn{4}{c}{not used (low quality)} \\
  \hline
\end{tabular}
\end{table*}

\subsection{PIONIER Observations}
HD~169142 was observed
with the VLTI/PIONIER instrument
on
one night in 2011 (ESO program \ProgramTwo)
and
three nights in 2013 (ESO program \ProgramOne).
PIONIER is a four-telescope beam combiner that works in the H band \citep{2011A&A...535A..67L},
providing spectrally dispersed
visibilities and
closure phases.
The observations were conducted using several baseline configurations%
\footnote{%
  \tiny%
    http://www.eso.org/sci/facilities/paranal/telescopes/vlti/configuration
}
of the 1.8~m auxiliary telescopes (ATs).
Each observation was preceded and followed by the observation of a calibrator
to characterize the instrumental transfer function.
A complete observation log is presented in Table \ref{table:observation.log}.
The data were reduced with the \texttt{PNDRS} package as described in \citet{2011A&A...535A..67L}.
The reduced archival OIFITS files were retrieved from
the Optical Interferometry Database \citep[OiDB,][]{2014SPIE.9146E..0OH}.

Out of the seven interferometric observations,
we analyzed the first six,
and discarded the last one (2011d)
because of its low data quality.
The $uv$-coverage of the final data set is plotted in Fig. \ref{fig:uv.coverage}.
The measured visibilities as function of wavelength are plotted in Fig. 4,
and the band-averaged visibilities are plotted in Fig. \ref{fig:GM}.
The measured closure phases are not plotted because they are all close to zero.
For the estimation of the uncertainty of the visibilities,
we assumed a calibration error of 5\%,
and quadratically added it
to the random error given by the reduction software.

\subsection{The optical-NIR SED}

For the interpretation of the interferometric data,
complementary photometric data are essential.
We adopted
the photometric observations obtained in the UBVRIJHK bands with the Sutherland SAAO telescope
\citep[performed in 2014,][]{2017A&A...599A..85L},
the optical photometry collected in \citet{2015ApJ...798...94W},
and the observations with SpeX spectrograph
 on NASA's Infrared Telescope Facility %(IRTF)
\citep[performed in 2013,][]{2015ApJ...798...94W}.
The compiled SED is shown in Fig. \ref{fig:SED}.
Among the NIR photometric data,
we adopted only those from epochs close to the interferometric observations,
bearing in mind the fact that HD~169142 is variable in NIR.
In the optical band we adopted all the available data
after confirming that there are no significant variability in the data set, which spans $\sim25$ years.
We assumed a $5\%$ uncertainty for each flux point.

\fi

\iftrue
\section{Geometric modeling}

In order to characterize the size of the NIR-emitting warm dust,
as a first step we performed a modeling of the emission with a simple ring geometry.
For this purpose, we first decomposed the SED to constrain
the fractional contribution of the flux originating from
the warm dust.
This decomposition also provides a rough estimation of the dust temperature.
In this section we first describe our model,
and then describe the process of SED de-composition and ring-fitting.
In this simple modeling,
we use only band-averaged visibilities.

In our model, the optical to NIR continuum of HD~169142 arises
from three origins,
the central star,
a sub-$\AU$ dust component,
and an extended halo.
Following \citet{2015ApJ...798...94W},
we modeled the central star as a Kurucz spectrum \citep{1992IAUS..149..225K}
with $T=7500~\Kelvin$, $\log g=4$,
and $A_v=0$.
The sub-AU warm dust radiates thermal emission in the NIR,
which we modeled as single-temperature blackbody.
Cold material at much larger distance,
which we generally refer to as the ``halo'',
scatters the light from the star and the warm dust,
but does not emit in the NIR by itself%
.
Therefore, the total flux of the whole system is
\begin{equation}
F(\lambda)
=
F_\mathrm{star}
+ F_\mathrm{dust}
+ F_\mathrm{halo}
=
F_\mathrm{star}
+ F_\mathrm{dust}
+ k(F_\mathrm{star}
+ F_\mathrm{dust})
,\end{equation}
where $k$ is the fraction of light scattered by the halo.
For simplicity, we assume the halo to have wavelength-independent albedo
and that it scatters the same fraction of the stellar light and dust continuum.

\subsection{SED decomposition}

The influence of the halo component to the SED is enhancing
the flux by a factor of $1+k$,
while not altering the shape.
Therefore, it is still possible to decompose the SED into
a hot photospheric component and a warm component.
The free parameters in the models are the bolometric flux
$F_\mathrm{h}$ of the hot Kurucz part, bolometric flux $F_\mathrm{w}$ of the warm part,
and the temperature of the dust $T_\mathrm{d}$.
We note that $F_\mathrm{h}$ includes
not only the flux directly from the central star,
but also the scattered stellar light from the halo.
Similarly, $F_\mathrm{w}$ includes both the contribution
directly from the dust and the photons that are radiated by dust and then scattered
by the halo.

We ran least-squares fitting to constrain the parameters.
The best-fit model and the observations are shown in Fig. \ref{fig:SED}.
The model parameters are listed in
Table \ref{table:SED.model}.
In this model, the fraction of flux contributed by the warm component
at H band
was $f_\mathrm{w}=22.3\pm2.1\%$.
This constraint was then used in the modeling of the interferometric data.

\begin{table}
\caption{
Results of our modeling of the SED of HD~169142 with
a Kurucz plus blackbody model.
The first three parameters are
free parameters in the model fitting;
the others are derived quantities.
The uncertainties are estimated with a bootstrapping method.
}
\label{table:SED.model}
\center
\begin{tabular}{ccccc}
\hline
\hline
Parameter & Best-fit value\\
\hline
$F_\mathrm{h}~[\mathrm{10^{-9}~erg~sec^{-1}~cm^{-2}}]$  & $14.6\pm0.24$\\
$F_\mathrm{w}~[\mathrm{10^{-9}~erg~sec^{-1}~cm^{-2}}]$  & $0.86\pm0.044$\\
$T_\mathrm{d}~[\Kelvin]$  & $1543\pm60$\\
\hline
$  \chi^2_\mathrm{red}$     & $3.16$\\
\hline
$f_\mathrm{w}~[\%]$      & $22.3\pm2.1$\\
\hline
\end{tabular}
\end{table}

\subsection{Ring model}
In the following we utilize our interferometric measurements
to extract information on the geometry of the inner disk.
We used a face-on model,
according to the constraints on inclination from previous observations.
High-resolution direct imaging of HD~169142 (summarized in Sect. 1)
has shown
that the outer disk is nearly face-on.
Modeling of $^{12}$CO line emission \citep{2006AJ....131.2290R}
gave an estimation of $i\approx13\degree$ for the outer disk.
Inclination of the inner disk has been estimated to be
$i\approx22\degree$
with PIONIER observations
\citep{2017A&A...599A..85L}.
All these observations indicate that the inner and outer disks
are nearly aligned to each other and both are nearly face-on.
Our model consists of the star, an infinitesimally narrow circular ring,
and an over-resolved halo component.
The total visibility is therefore
\begin{equation}
  V =
  f_\mathrm{star} V_\mathrm{star}
  +
  f_\mathrm{ring} V_\mathrm{ring}
  +
  f_\mathrm{halo} V_\mathrm{halo}
,\end{equation}
where $f_\mathrm{star}$, $f_\mathrm{ring}$, and $f_\mathrm{halo}$
are the fractions of flux contributed by each component.
The fractions follow the rule $f_\mathrm{star} + f_\mathrm{ring} + f_\mathrm{halo} \equiv 1$.
We assumed that the star is completely unresolved
and the halo is completely overresolved,
so that $V_\mathrm{star}\equiv1$,
and $V_\mathrm{halo}\equiv0$.
Measuring at wavelength $\lambda$, with a baseline of length $B$,
the visibility of the circular ring is the Bessel function
\begin{equation}
V_\mathrm{ring} =
J_0(2\pi \nu r_\mathrm{ring})
,\end{equation}
where $r_\mathrm{ring}$ is the angular ring radius,
and $\nu=B/\lambda$ is the spatial frequency.
$f_\mathrm{halo}$ and $r_\mathrm{ring}$ are free model parameters.
The other two flux fractions are derived as
\begin{equation}
f_\mathrm{ring} = f_\mathrm{w}(1-f_\mathrm{halo})
\end{equation}
and
\begin{equation}
  f_\mathrm{star} = (1-f_\mathrm{w}) (1-f_\mathrm{halo})
,\end{equation}
where $f_\mathrm{w}$ is constrained by the SED decomposition (see Sect. 3.1).
Our best-fit model and error estimations are presented
in Table \ref{table:GM} and Fig. \ref{fig:GM}.
The uncertainties are estimated with a bootstrapping method.
In each of our bootstrapping samples, both the fluxes and the visibilities
are re-sampled.
Therefore, the error estimations account not only for the uncertainties in visibilitiy measurement,
but also for the uncertainties in the photometry.

In the best-fit model, the ring radius is ${\sim}0.08~\AU$,
which is in rough agreement with the dust sublimation radius assuming $T_\mathrm{sub}\sim1500~\Kelvin$.
The fraction between bolometric flux of the warm and the photospheric component is
$F_\mathrm{w}/F_\mathrm{h}\sim0.06$.
Therefore, the model is consistent with
warm dust absorbing
${\sim}6\%$ of all the stellar light
and re-emiting the energy in the NIR.

A halo component is clearly needed to account for the fact that
the visibility is already lower than $1.0$ at the shortest baselines ($B\sim10~\meter$).
In the best-fit model the flux fraction of the halo
is ${\sim}12\%$,
only slightly higher than the estimate of ${\sim}8\%$ in \citet{2017A&A...599A..85L}.
Halo components,
representing scattered light from extended ($>1~\AU$) cold material,
are frequently detected in NIR interferometric observations of Herbig stars,
with typical flux ratios of $5\%\enDash20\%$
\citep[e.g.,][]{%
2006ApJ...647..444M%
,2012A&A...541A.104C%
,2016A&A...590A..96K%
}.
Plausible origins
of the halo material include an infalling remnant envelope,
dust entrained in the stellar wind or outflow \citep{2006ApJ...647..444M},
or a flaring outer disk that scatters the stellar light \citep{2008ApJ...673L..63P}.
In the particular case of HD~169142, stellar light scattered from the outer disk
has been detected in direct imaging,
and such imaging has indeed been employed to reveal the complex disk structure
\citep{2013ApJ...766L...2Q}.
The outer disk's scale height of ${\sim}0.1$ \citep{2017ApJ...838...20M}
at $R\sim30~\AU$
% \footnote{scaled to $D=117\pc$.}
also indicates its capability of scattering ${\sim}10\%$ of stellar light.
Therefore,
it is likely that the outer disk is the main contributor to the halo light in HD~169142.

To summarize, our {\GM}ing indicates that
the NIR brightness distribution in HD~169142 can be well interpreted with
a central star, passively heated warm dust
circular-symmetrically distributed around the star,
and an extended scattering component.
Therefore, in the next step we set up our {\RT} models within this frame,
and perform numerical simulation
in order to further constrain the properties of the warm dust.

\begin{table}
\caption{
Results of our modeling of the visibilities of HD~169142 with
star-ring-halo model.
The first two parameters are
free parameters in the model fitting;
the others are derived quantities.
The uncertainties are estimated with a bootstrapping method.
}
\label{table:GM}
\center
\begin{tabular}{ccccc}
\hline
\hline
Parameter & Best-fit value\\
\hline
$r_\mathrm{ring}~[\mas]$    & $0.655\pm0.061$\\
$f_\mathrm{halo}~[\%]$      & $11.9\pm0.6$\\
\hline
$  \chi^2_\mathrm{red}$     & $0.91$\\
\hline
$f_\mathrm{star}~[\%]$      & $68.5\pm1.9$\\
$f_\mathrm{ring}~[\%]$      & $19.6\pm1.9$\\
$L_\mathrm{star}~[L_\sun~\left(\frac{d}{117~\pc}\right)^2]$  & $5.49\pm0.10$\\
$R_\mathrm{ring}~[\AU~\left(\frac{d}{117~\pc}\right)]$     & $0.077\pm0.007$\\
\hline
\end{tabular}
\end{table}

\fi

\iftrue

\newcommand{\TabularParameterDescription}{
  The listed parameters are:\newline
  $L_*=\textrm{stellar luminosity }$,\newline
  $i=\textrm{inclination of the disk}$,\newline
  $\mathrm{PA}=\textrm{positional angle of rotation axis the disk}$,\newline
  and the following parameters for the inner sub-AU disk,\newline
  $R_{\mathrm{in}}=\textrm{inner radius of the inner disk }$,\newline
  $W_\mathrm{ring}=(R_\mathrm{out}-R_\mathrm{in}) / R_\mathrm{in}$
  (relative radial width of the inner component),\newline
  $R_{\mathrm{out}}=\textrm{outer radius of the inner disk }$,\newline
  $h_{\mathrm{in}}=\textrm{scale height of the inner disk at its inner radius}$,\newline
  $q=\textrm{scale-height power-law index of the inner disk}$,\newline
  $p =\textrm{surface-density power-law index}$,\newline
  $\tau = \textrm{Planck-averaged midplane optical depth at }7500~\Kelvin$,\newline
  $n = \textrm{power-law index of grain size distribution}$,\newline
  $f_\mathrm{carbon} = \textrm{fraction of carbon in the dust}$.\newline
  %$M_{\mathrm{dust}} = \textrm{dust mass}$
}
\newcommand{\TabularContent}{Tabular: undefined!}

\renewcommand{\TabularContent}{ 
  \begin{tabular}{ccccc}
    \hline
    \hline
    \HeadLineForModelTabular
    \\
    \hline
    \TabulatedCodeForParameterSystem
    \hline
    \multicolumn{4}{l}{
      Component 1 (inner disk)
    }\\
    \TabulatedCodeForParameterInnerDisk
    \hline
    \multicolumn{4}{l}{
      Component 2 (halo)
    }\\
    \TabulatedCodeForParameterOuterDisk
    \hline
    $\chi^2_\mathrm{red}$
    &
    &
    & \BestChiSquare
    \\
    \hline
  \end{tabular}
}

% This part defines the two "version"s of the model figure caption.
\newcommand{\ModelFigureCaptionTextOne}{
  Model scanning run {\RunNameShort} (\ModelClassDescription),
  and the model parameters of the best model {\ModelNameShort}
  with minimum $\chi^2_\mathrm{red}$,
  which is calculated by comparing each model with the data set.

  {\it a:}
  Parameter ranges of model scanning run {\RunNameShort},
  number of tested parameter values (steps) per parameter,
  and parameters of the best model {\ModelNameShort}.
  {
    \TabularParameterDescription
  }

  {\it b:}
  $\chi^2_\mathrm{red}$ maps of Model run \RunName.
  {
    For each subset of parameters, the $\chi^2_\mathrm{red}$
    shown is the lowest value for all combinations with other paramters.
    For example, for each pair of ($R_\mathrm{in}$, $\tau$) values,
    the $\chi^2_\mathrm{red}$ values for all possible
    ($L_*$, $n$, $f_\mathrm{carbon}$) combinations within the described ranges
    were compared
    and the minimum value found
    is plotted into the map at left panel.
  }

  {\it c:} SED of HD~169142.
  The lines denote the model contributions from different radial regions.
  The red dots are the observations.
  $5\%$ uncertainty is assumed for each observation.

  {\it d:} NIR visibilities %, and closure phases CP
  (red dots: observations; black lines: model)
  from our VLTI/PIONIER observations.

}
\newcommand{\ModelFigureCaptionTextTwo}{
  As Fig.~4, but for model run {\RunName}, compared with the data set.
  Model {\ModelNameShort} is the model with minimum $\chi^2_\mathrm{red}$.
  {
  %  \TabularParameterDescription
  }
}

\newcounter{CounterModel}
\newboolean{FirstModel}
\newcommand{\RunNameShort}{}
\newcommand{\RunNameLong}{}
\newcommand{\RunName}{\RunNameShort}
\newcommand{\DatasetName}{}
\newcommand{\ModelNameLong}{}
\newcommand{\ModelNameShort}{\RunName\DatasetName}
\newcommand{\ModelClassDescription}{}
\newcommand{\ModelName}{\ModelNameLong}
\newcommand{\ModelFigureCaptionText}{}
\newcommand{\TableName}{}
\newcommand{\BestChiSquare}{$Undefined??$}

  \setlength\fboxsep{0pt}
  \setlength\fboxrule{0.1pt}
\newcommand{\SubFig}[2]{
  % #1 is the index
  % #2 is the content
  \mbox{
    \parbox[t]{0cm}{
      \vspace{0.1cm}
      \mbox{\large \bf
        {#1}%\hspace{9mm}
      }
    }
    \nolinebreak
    \parbox[t]{1.0\ColumnWidth}{%
      %\rule{0mm}{0mm}\newline
      \vspace{0mm}
      \parbox[b]{1.0\ColumnWidth}{#2}
    }
  }
}

%
% The following defines the ways to plot the information for the model (scanning).
\newcommand\StandardTableAndFigure{}

\renewcommand\StandardTableAndFigure{

  \stepcounter{CounterModel}
  \ifthenelse{\arabic{CounterModel}=1}{\setboolean{FirstModel}{true}}%
  {\setboolean{FirstModel}{false}}

  \ifthenelse{\boolean{FirstModel}}{
    \renewcommand{\ModelFigureCaptionText}{\ModelFigureCaptionTextOne}
  }{
    \renewcommand{\ModelFigureCaptionText}{\ModelFigureCaptionTextTwo}
  }

  \begin{figure*}
    \begin{minipage}[t]{2.0\ColumnWidth}
      \begin{minipage}[t]{\ColumnWidth}
        \SubFig{a}{
          \parbox{\ColumnWidth}{ % the part for the table
            \setlength{\tabcolsep}{9pt}
            \renewcommand{\arraystretch}{1.2}
            \hspace{0.5cm}
            \parbox{0.8\ColumnWidth}{
        \setlength{\baselineskip}{10pt}
  \ifthenelse{\boolean{RefereeMode}}{
        \renewcommand{\arraystretch}{0.9}
  }{
  }
              \TabularContent
            }
          }
        }

        \SubFig{b}{
          \includegraphics[width=1.0\ColumnWidth,angle=-0]{fig/\ModelName/chi2.eps}
        }

      \end{minipage}
      \begin{minipage}[t]{1.0\ColumnWidth}

        \SubFig{c}{
          \mbox{\includegraphics[scale=0.25,angle=-0,trim=0 0 -0.0cm 0]{fig/\ModelName/SED.eps}}
        }

        \SubFig{d}{
          \parbox{\ColumnWidth}{
            \includegraphics[width=1.0\ColumnWidth,angle=-0]{fig/\ModelName/V.NIR.eps}
          }
        }

      \end{minipage}
    \end{minipage}
  \ifthenelse{\boolean{RefereeMode}}{
    \caption{
        \setlength{\baselineskip}{12pt}
    \ModelFigureCaptionText }
  }{
    \caption{
    \ModelFigureCaptionText }
  }
    \label{figure:\ModelName}
  \end{figure*}

}

\newcommand\FigRef{%
  \ref{figure:MCRT_\RunNameLong_dataset\DatasetName}%
}
\newcommand\StandardTableAndFigureList{%
  Fig.~\FigRef%
}

\newcommand{\ModelNumber}{"Model Number"}
\newcommand{\HeadLineForModelTabular}{%
  Parameter                 & Parameter range                   & Steps     & Model \ModelNameShort
}

% For each scan (AA, BA, ...), redefine the relevant parameters, figures, etc.
% Each "\scanXX" will only do setting, and never insert any real content.

% Note that, often, some values need to be "recomputed".

\newcommand{\RecomputerScan}{
    \renewcommand{\TableName}{table:parameters:\RunNameLong_dataset\DatasetName}
  \renewcommand{\ModelName}{MCRT_\RunNameLong_dataset\DatasetName}
  % read in a file which defines
    % 1. the descriptions of the parameters.

\providecommand\TabulatedCodeForParameterSystem{}
\renewcommand\TabulatedCodeForParameterSystem{
$L_*~[L_\sun]$
  & \ParameterRange{5.5}{4}
  & $4$
  & $5$\\
$i~[\degree]$
  & $0$
  & $-$
  & $0$\\
$\mathrm{PA}~[\degree]$
  & $0$
  & $-$
  & $0$\\
}

\providecommand\TabulatedCodeForParameterInnerDisk{}
\renewcommand\TabulatedCodeForParameterInnerDisk{
$R_\mathrm{in}$ $[\AU]$
  & \ParameterRange{0.04}{0.4}
  & $8$
  & $0.077$\\
$W_\mathrm{ring}$
  & $0.5$
  & $-$
  & $0.5$\\
$h_\mathrm{in}$
  & \ParameterRange{0.01}{0.1}
  & $6$
  & $0.025$\\
$q$
  & $0$
  & $-$
  & $0$\\
$p$
  & $0$
  & $-$
  & $0$\\
$\tau$
  & $1000$
  & $-$
  & $1000$\\
$n$
  & \ParameterRange{-5}{-2}
  & $7$
  & $-3.5$\\
$f_\mathrm{carbon}$
  & $0.5$
  & $-$
  & $0.5$\\
}

\providecommand\ModelNumber{}
\renewcommand\ModelNumber{$ 4{\times}8{\times}6{\times}7 = 1344 $
}

\providecommand\BestChiSquare{}
\renewcommand\BestChiSquare{1.35
}

\providecommand\TabulatedCodeForParameterOuterDisk{}
\renewcommand\TabulatedCodeForParameterOuterDisk{
$k$
& $0.13$
& $-$
& $0.13$
\\
}

}

\newcommand{\SetToScanAA}{
  \renewcommand{\RunNameLong}{1comp_Thin}
  \renewcommand{\RunNameShort}{A}
  \renewcommand{\DatasetName}{1}
  \renewcommand{\ModelClassDescription}{optically thin inner disk}

    \renewcommand{\TableName}{table:parameters:\RunNameLong_dataset\DatasetName}
  \renewcommand{\ModelName}{MCRT_\RunNameLong_dataset\DatasetName}
  % read in a file which defines
    % 1. the descriptions of the parameters.

\providecommand\TabulatedCodeForParameterSystem{}
\renewcommand\TabulatedCodeForParameterSystem{
$L_*~[L_\sun]$
  & \ParameterRange{5.5}{4}
  & $4$
  & $5$\\
$i~[\degree]$
  & $0$
  & $-$
  & $0$\\
$\mathrm{PA}~[\degree]$
  & $0$
  & $-$
  & $0$\\
}

\providecommand\TabulatedCodeForParameterInnerDisk{}
\renewcommand\TabulatedCodeForParameterInnerDisk{
$R_\mathrm{in}$ $[\AU]$
  & \ParameterRange{0.04}{0.4}
  & $8$
  & $0.077$\\
$W_\mathrm{ring}$
  & $0.5$
  & $-$
  & $0.5$\\
$h_\mathrm{in}$
  & \ParameterRange{0.01}{0.1}
  & $6$
  & $0.025$\\
$q$
  & $0$
  & $-$
  & $0$\\
$p$
  & $0$
  & $-$
  & $0$\\
$\tau$
  & $1000$
  & $-$
  & $1000$\\
$n$
  & \ParameterRange{-5}{-2}
  & $7$
  & $-3.5$\\
$f_\mathrm{carbon}$
  & $0.5$
  & $-$
  & $0.5$\\
}

\providecommand\ModelNumber{}
\renewcommand\ModelNumber{$ 4{\times}8{\times}6{\times}7 = 1344 $
}

\providecommand\BestChiSquare{}
\renewcommand\BestChiSquare{1.35
}

\providecommand\TabulatedCodeForParameterOuterDisk{}
\renewcommand\TabulatedCodeForParameterOuterDisk{
$k$
& $0.13$
& $-$
& $0.13$
\\
}

}

\newcommand{\SetToScanAAF}{
  \renewcommand{\RunNameLong}{1comp_Thin_fine}
  \renewcommand{\RunNameShort}{AF}
  \renewcommand{\DatasetName}{1}
  \renewcommand{\ModelClassDescription}{optically thin inner disk}

    \renewcommand{\TableName}{table:parameters:\RunNameLong_dataset\DatasetName}
  \renewcommand{\ModelName}{MCRT_\RunNameLong_dataset\DatasetName}
  % read in a file which defines
    % 1. the descriptions of the parameters.

}

\newcommand{\SetToScanAW}{
  \renewcommand{\RunNameLong}{1comp_Thin_scanW}
  \renewcommand{\RunNameShort}{AW}
  \renewcommand{\DatasetName}{1}
  \renewcommand{\ModelClassDescription}{optically thin inner disk}

    \renewcommand{\TableName}{table:parameters:\RunNameLong_dataset\DatasetName}
  \renewcommand{\ModelName}{MCRT_\RunNameLong_dataset\DatasetName}
  % read in a file which defines
    % 1. the descriptions of the parameters.

}

\newcommand{\SetToScanBA}{
  \renewcommand{\RunNameLong}{1comp_Thick}
  \renewcommand{\RunNameShort}{B}
  \renewcommand{\DatasetName}{1}
  \renewcommand{\ModelClassDescription}{optically thick inner disk}

    \renewcommand{\TableName}{table:parameters:\RunNameLong_dataset\DatasetName}
  \renewcommand{\ModelName}{MCRT_\RunNameLong_dataset\DatasetName}
  % read in a file which defines
    % 1. the descriptions of the parameters.

}

\newcommand{\SetToScanBAF}{
  \renewcommand{\RunNameLong}{1comp_Thick_fine}
  \renewcommand{\RunNameShort}{BF}
  \renewcommand{\DatasetName}{1}
  \renewcommand{\ModelClassDescription}{optically thick inner disk}

    \renewcommand{\TableName}{table:parameters:\RunNameLong_dataset\DatasetName}
  \renewcommand{\ModelName}{MCRT_\RunNameLong_dataset\DatasetName}
  % read in a file which defines
    % 1. the descriptions of the parameters.

}

\section{{\MCRT} modeling of the interferometric data and the SED}
% An summary of the model setup.
In this section we present a more physical modeling approach.
In our {\RT} modeling,
we have tried to reproduce the interferometric data and the SED simultaneously in the optical-NIR
with a disk model
consisting of a sub-$\AU$ inner disk and a scattering halo component.

For the dust disk,
we assumed a power-law radial distribution for its surface density $\Sigma$,
\begin{equation}
\Sigma\left(r\right) = \Sigma_\mathrm{in}\left(\frac{r}{R_\mathrm{in}}\right)^{p}
,~
R_\mathrm{in} < r < R_\mathrm{out}
, \end{equation}
where $\Sigma_\mathrm{in}$ is the surface density at the inner radius $R_\mathrm{in}$,
$p$ is the power-law exponent,
and $R_\mathrm{out}$ is the outer radius of the dust disk.
We assumed the vertical dust density distribution to be a Gaussian function,
and the resulting density structure of the disk is
\begin{equation}
\rho \left(r,z\right) = \Sigma\left(r\right)\frac{1}{\sqrt{2\pi}H} \exp \left(-\frac{z^2}{2 H^2} \right),
\end{equation}
where $\rho(r,z)$ denotes the dust density as a function of $r$ and the height $z$ above the mid-plane,
and $H$ is the scale height.
We assumed that the dependence of $H$ on $r$ is also a power-law,  
\begin{equation}
h\left(r\right) \equiv \frac{H\left(r\right)}{r} = h_\mathrm{in}\left(\frac{r}{R_\mathrm{in}}\right)^{q},
\end{equation}
where $h\left(r\right)$ is the dimensionless scale height,
$h_\mathrm{in}$ is the dimensionless scale height at the inner radius,
and $q$ is the power-law index.

The main purpose of this paper is to constrain
the properties of the inner disk,
including its size, geometry, and dust composition
with our new NIR interferometric data.
Therefore, we tested two types of models for the inner sub-$\AU$ region:
optically thin dust with a large scale height (model type A),
or optically thick dust with a flat geometry (model type B).
We used a mixture of amorphous carbon \citep{1998A&A...332..291J}
and astronomical silicate \citep{1984ApJ...285...89D},
and tested different dust compositions by adjusting the
parameter $f_\mathrm{carbon}$ (fraction of carbon).
We assumed a power-law grain size distribution
\begin{equation}
n(a)\propto a^{n}, a_\mathrm{min}<a<a_\mathrm{max}
.\end{equation}
With the limited information available,
it is impossible to constrain the parameters $a_\mathrm{min}$,
$a_\mathrm{max}$, and $n$ simultaneously.
We chose to use fixed size limits $a_\mathrm{min}=0.01~\mum$
and $a_\mathrm{max}=10^3~\mum$,
and used the power-law index $n$ to control the fraction
of sub-mm and larger grains.

We tested different stellar luminosities
around the value estimated in Sect. 3.2.
A simplification in the {\GM}ing is that the scattering of stellar light
from the inner dust is neglected.
When this process is fully considered in our {\RT} modeling,
the best-fit luminosity value is found to be slightly lower.

For each set of stellar and inner disk parameters,
we performed {\MCRT} (MCRT) simulation with RADMC3D,
rendering the spatial and spectral distribution of the radiation.
The scattered light from larger-scale material
(corresponding to the ``halo'' component in the \GM)
is not directly included in the MCRT simulation,
but taken into consideration in a parametrized way.
We assumed the halo to scatter the light from the central star
and the inner disk with a wavelength-independent fraction $k$.
Therefore, the effects of the halo are that
it increases the total flux from the system by a factor of $1+k$,
and reduces the visibilities by the same factor.
Therefore, after each MCRT run,
the final SED and $V^2$ can be analytically calculated.
We used a fixed $k=13\%$, which is derived from
$f_\mathrm{halo}=\frac{k}{1+k}=12\%$ in Sect. 3.2.

\subsection{Fitting with model type A (optically thin inner disk)}

In the optically thin case, the amount of stellar light 
that the dust absorbs and re-emits in the NIR
is roughly proportional to the optical depth $\tau$,
and to the scale height $h_\mathrm{in}$ of the dust.
Obviously there is a degeneracy between the two parameters.
We chose to use a fixed $h_\mathrm{in}=0.6$, and use $\tau$ as a proxy to the amount of NIR excess.

While our interferometric data set constrained the size of the inner NIR-emitting region,
it did not contain enough information to constrain the radial dust distribution in detail,
because the inner dust
is only partially resolved
even with the longest baselines.
This ambiguity is reflected in a degenaracy between the relative width
$W_\mathrm{ring}= ( R_\mathrm{out}-R_\mathrm{in} ) / R_\mathrm{in}$
and the inner radius $R_\mathrm{in}$,
as demonstrated in Appendix A.2.
We therefore chose to use a fixed relative width
 $W_\mathrm{ring} = 0.5$
and use $R_\mathrm{in}$ to characterize the disk size.

\SetToScanAAF%
\StandardTableAndFigure
We tested ${\sim}7000$ models in order to find a model of type A
that can approximately
reproduce the NIR visibilities and the NIR SED.
The best-fit model {\ModelNameShort} is shown in
\StandardTableAndFigureList.
We present in Appendix A.1 a description of the parameter scanning process,
in which we start the scanning from broad parameter ranges
and then narrow down to a finer parameter grid.

We find that
both the inner disk radius and the dust grain size distribution
are well constrained by the modeling.
In the best-fit model,
the inner disk has an $R_\mathrm{in}$ of $0.07~\AU$,
and a power-law index $n$ of $-2$ for the grain size distribution.
Fitting of similar quality can be achieved for any distribution with $n\ge-3$.
Models with $n\le-3.5$ have much higher $\chi^2$ and are exluded.
The power law with $n\ge-3$  means that the dust consists mainly of
large grains with $a>1~\mum$.

In the following we discuss the reason why the modeling is sensitive to the parameter $n$.
The radiative-equilibrium temperature of optically thin dust irradiated
by a star with luminosity $L_*$ at a distance of $R$ is \citep[see, e.g.,][]{2010ARA&A..48..205D}
\begin{equation}
T =
\left(
\frac{L_*}
{ 16\pi \sigma \epsilon R^2 }
\right) ^{1/4}
,\end{equation}
where the cooling parameter $\epsilon$
is the ratio between the absorption coefficient of the dust in the NIR
and that in the optical.
$\sigma$ is the Stefan-Boltzmann constant.
Therefore, for a given $R$ (constrained with the interferometric observations),
the color of the NIR dust emission is sensitive to $\epsilon$,
which in turn depends on the grain size distribution.
For example,
dust with 
$a_\mathrm{min}=0.01~\mum$,
$a_\mathrm{max}=10^3~\mum$,
and $n=-5$
has $\epsilon \approx 0.2$,
which causes an increase of temperature by a factor of $\epsilon^{-1/4}\approx1.5$
compared to gray dust.
Such increase will have a strong effect on the model SED.

For grain sizes larger than $1~\mum$, the dust will have a flat opacity curve
in the optical-NIR range,
and therefore $\epsilon\sim1$,
independent of chemical composition
\citep[see, e.g., Fig. 8 in][]{2010ARA&A..48..205D}.
Therefore, our modeling sets no constraint on $f_\mathrm{carbon}$.
We cannot get complementary constraints on the chemical composition
from spectroscopic study around the $10~\mum$ silicate feature,
because the N-band emission in HD~169142 is dominated by strong PAH features
\citep{2014A&A...563A..78M}.

In this optically thin case,
the dust temperature decreases smoothly with the distance from the central star,
$T \propto r^{-1/2}$.
For the narrow radial width $W_\mathrm{ring}=0.5$ that we assumed,
the range of temperatures is also narrow.
In the best-fit model,
the dust temperature ranges between
${\sim}1300~\Kelvin$ at the outer edge
and
${\sim}1600~\Kelvin$ at the inner edge.

\subsection{Fitting with model type B (optically thick inner disk)}

In the optically thick case, the amount of stellar light 
that the dust absorbs and re-emits in the NIR
becomes insensitive to the optical depth $\tau$ of the dust,
and depends mainly on the scale height $h_\mathrm{in}$ of the dust.
We therefore use a fixed $\tau=10^3$, and use $h_\mathrm{in}$ to adjust the amount of NIR excess.

\SetToScanBAF%
\StandardTableAndFigure
We tested ${\sim}2000$ models in order to find a model of type B
that can approximately
reproduce the NIR visibilities and the NIR SED.
The best-fit model {\ModelNameShort} is shown in
\StandardTableAndFigureList.
The parameter scanning process is described in Appendix A.2.

While the size of the inner rim is tightly constrained to be ${\sim}0.06~\AU$,
the dust grain size distribution is not constrained at all.
The temperature distribution of passively heated optically thick dust is highly inhomogeneous,
with hot surface layers covering the cold interior region.
The color of the overall emission is determined by complicated {\RT} processes
involving both the surface layers and the interior.
Numerical studies of the {\RT} processes show that the overall color is insensitive to $\epsilon$
\citep[see e.g.,][]{2010ARA&A..48..205D,2016A&A...586A..54C}.

The hottest part in the best-fit model is the surface layer of the inner rim.
This layer is
directly irradiated by the central star
and has a temperature of ${\sim}2400~\Kelvin$.
We found that this surface temperature is also insensitive to $n$ or $\epsilon$,
in the $\epsilon$ range we tested ($0.2\lesssim\epsilon\lesssim1$).
The surface temperature is higher than the sublimation temperature
of silicate or carbonaceous material.
Therefore, to establish a self-consistent model,
refractory dust species that can exist at ${\sim}2400~\Kelvin$ should be added to the model BF1.
We did not perform {\RT} computing for a model including refractory dust,
because our modeling already shows that neither the surface temperature
nor the modeled observables
are sensitive to the choice of species.
Therefore, our conclusion is that
the observations can be reproduced by an optically thick dust,
with inner rim radius ${\sim}0.06~\AU$ and scale height $H/R \sim 0.03$,
consisting of a mixture of refractory dust and normal dust (silicate and carbonaceous material).

\fi

\section{Discussion}
\iftrue

\subsection
{Constraints on dust properties}

The inner sub-AU disk of HD~169142 is likely in rapid variation \citep{2015ApJ...798...94W}.
The interferometric PIONIER observations allow us to
study its structure and dust properties around the years 2011-2013.
Our {\GM}ing shows that its NIR excess arises from warm dust (${\sim}1500~\Kelvin$)
at a distance of ${\sim}0.08~\AU$
from the central star,
consistent with dust sublimation radius for optically thin gray dust.
With {\RT} modeling we found that
the observations are consistent with optically thin gray dust
lying at $R_\mathrm{in} \sim 0.07~\AU$.
In the best-fit model of this type (model AF1),
the temperature of the dust is between ${\sim}1300~\Kelvin$ and ${\sim}1600~\Kelvin$.
Therefore, both the location and temperature of dust in model AF1 is similar
to that estimated in {\GM}ing.
Models with sub-micron optically thin dust are excluded
because such dust will have much higher temperature at similar distance.

A model with optically thick dust with inner rim at $R_\mathrm{in} \sim 0.06~\AU$
can also reproduce the observations.
%will radiate with similar ``color" and cannot be excluded.
However, due to the highly inhomogeneous temperature distribution,
with $T\sim2400~\Kelvin$ in its hottest part,
this model is plausible only if refractory dust species exist in the inner disk.
Evidence for the existence of refractory dust has been found in other Herbig stars.
For example, interferometric study of HD~163296 \citep{2010A&A...511A..74B}
indicates inner refractory dust with a temperature of $2100\enDash2300~\Kelvin$.

Because models with either optically thin or
thick disks can reproduce the data,
it is obvious that a hybrid model
consisting of both optically thin and thick components
is also possible,
and the fraction of NIR light contributed by each part is unconstrained.
In such a model, gray dust is again preferred to sub-micron dust.

The inner disk size in each of our models is significantly smaller than the size
of $R_\mathrm{in}\sim0.19~\AU$%
\footnote{scaled to $D=117\pc$.}
in the previous model in \citet{2015ApJ...798...94W}.
The fact that their model also reproduced the NIR SED reflects
a degeneracy between the radius $R_\mathrm{in}$ and the cooling parameter $\epsilon$ in SED modeling.
While increasing $R_\mathrm{in}$ will decrease the dust temperature and make the dust emission redder,
using dust with lower $\epsilon$ will increase the temperature of optically thin dust and make the emission bluer,
balancing the effect of increasing $R_\mathrm{in}$.
In the present work,
the interferometric observations constrain $R_\mathrm{in}$
and therefore break the degeneracy.

\subsection
{NIR variability of the inner disk}

In the scenario proposed by \citet{2015ApJ...798...94W},
the NIR variability in HD~169142
is caused
by changes in the dust distribution of an optically thin component,
while the optically thick core in their models
remains constant.
They proposed three models for the pre-2000 state of HD~169142,
in which the optically thin components differ from
the post-2000 state in scale height, radial width, and/or dust density.
Despite the differences between these models,
a common character in them is that the optically thin dust always has higher
mass than in the post-2000 state
by a factor of at least two,
indicating that this component is dissipating on a short timescale of ${\sim}10~$yr.
Their argument against the variation caused by varying height of an optically thick dust
is that
such a structural change,
while changing the ability of the dust in reprocessing stellar light into NIR emmision,
will also cause varying shadowing on the outer disk
and lead to variations in the emission from the outer disk.
The lack of MIR-FIR variation in HD~169142 \citep{2012ApJS..201...11K,2015ApJ...798...94W}
therefore may disfavor a varying optically thick dust.

If the NIR variation in HD~169142 is solely due to variation in the optically thin component,
then our two RT models for the inner dust in low state
will correspond to two scenarios (Fig. \ref{fig:sketch}) for the readjusting of the inner dust.
In scenario 1, the inner dust is optically thin
in the pre-2000 state
and has lost part of its mass in the post-2000 state (model AF1).
In scenario 2, the inner dust consists of an optically thin component and an optically thick core in the pre-2000 state.
The optically thin component has completely dissipated after 2000 (model BF1),
while the optically thick core remains constant.
In both scenarios, the NIR variabilities arise solely from changes affecting the optically thin component.

\begin{figure}
  \includegraphics[width=9cm]{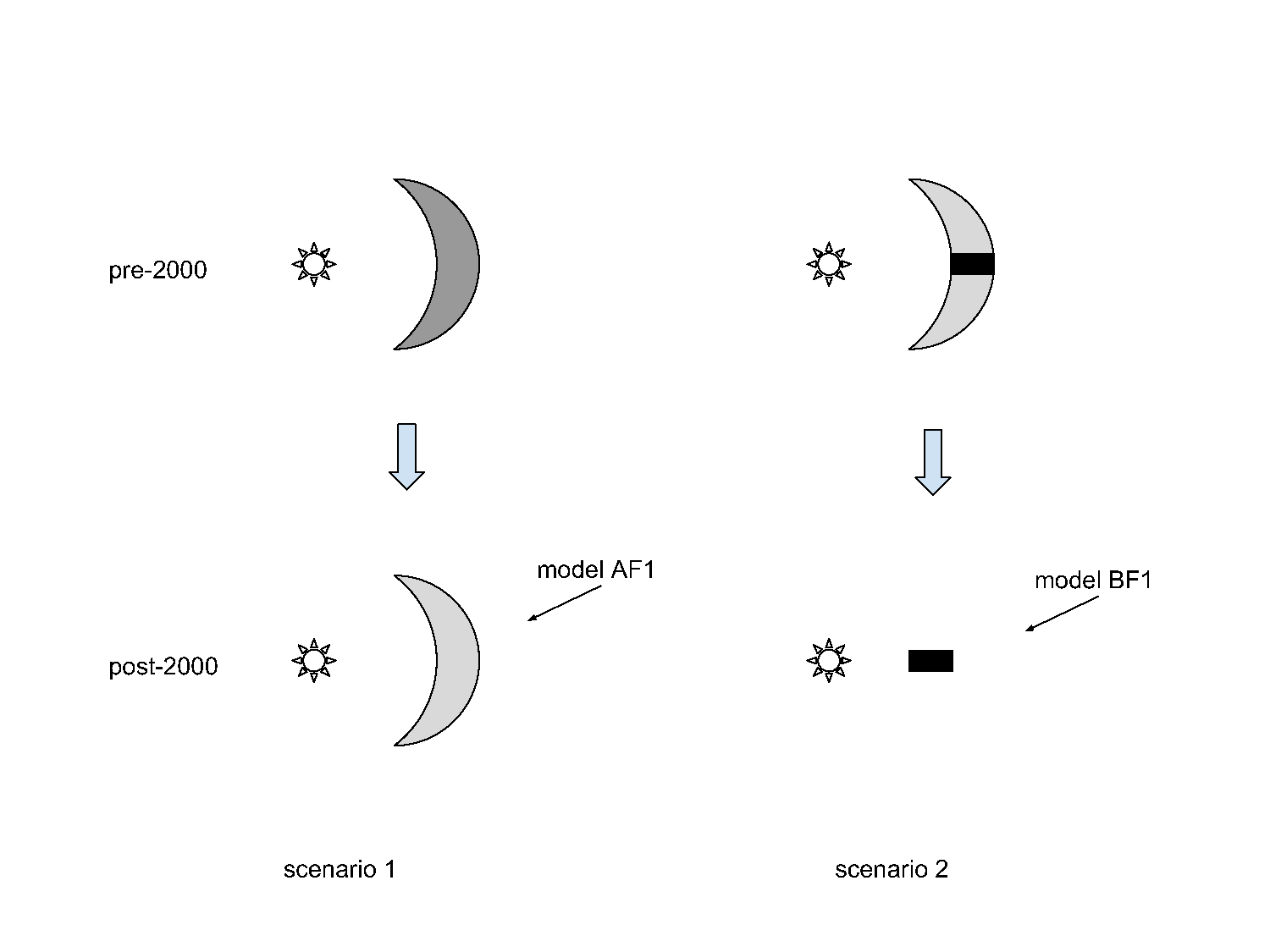}
  \caption{
    Sketch of the scenarios for the evolution of the inner sub-AU region in HD~169142.
  }
  \label{fig:sketch}
\end{figure}

The inner disk will probably continue varying in the following decades.
Following scenario 1,
if the inner optically thin dust continues to lose its mass
with the current mass-loss rate,
then the disk will evolve into a transitional disk within only several decades,
and this would be an extremely interesting case of disk evolution.
However, such a timescale
is much shorter than
the expected evolutional timescale of ${\sim}10^5~$yr
for disks in the pre-transitional and transitional phases.
It is also surprising
from statistical point of view
that an object is caught in this very short
evolution phase which lasts for less than 100 years,
considering the small number of Herbig stars observed.
Therefore, it is more likely
that the fading observed in the past decades
is only one episode in a series of fluctuations in dust amount and in its emission.
This can happen if the dust is being replenished in a time-dependent process.
Collisions between larger bodies can
provide secondary dust 
to the inner region of pre-transitional objects
\citep{%
2009ApJ...699.1822G,%
2006ApJ...637L.133E,%
2011A&A...531A..80K%
}.
Such replenishing can be occasionally enhanced by
a collision between the largest parental bodies
or by gravitational stirring from a nearby planet
\citep{2010RAA....10..383K}.
%Due to the short lifetime of the secondary dust
%of at most several decades \cite{Rhee2008},
%fluctuations on timescale of ${\sim}10~\mathrm{yr}$
%can be caused by the episodic replenishing.
%The fading timescale after each enhancing event
%would be determined by the lifetime of the released secondary dust,
%which is expected to be at most several decades \citep{2008ApJ...675..777R}.
%
Small amounts of dust from the outer disk
passing through the planet-induced gap region \citep{2016A&A...585A..35P}
can also be an unsteady supply of inner NIR-emitting dust
and cause NIR variations on timescale of several decades
\citep[e.g., in the case of GW~Ori,][]{2014A&A...570A.118F}.

For scenario 2, the optically thin dust
may be replenished by disk wind
or by material lifted up from
the optically thick disk by turbulent dynamic processes,
and these unsteady processes can cause variability
on timescale of months or years
\citep{2009ApJ...706L.168B}.
Another possible cause of future NIR variability
is changing of height of the optically thick component,
though this does not seem to have happened in the past decades,
as argued by \citet{2015ApJ...798...94W}.

Further NIR photometric observations,
which constrain the amount and temperature of the hot dust,
and NIR interferometry,
which constrains the radial location and grain size distribution of the dust,
will provide useful information on the nature of variabilities in the inner region.
Photometry at MIR-FIR wavelengths will be useful
in revealing the varying shadowing on the outer disk,
in case there will be any change in the height
of the inner optically thick component.

\subsection{Correlation between dust grain size and gap size?}

Statistical studies
on how the dust grain size distribution depends on other disk parameters
can provide useful constraint
on theoretical modeling of disk evolution.
In the following we compare our results with
inner dust properties in several other Herbig stars with pre-transitional disks.

HD~100546 has a gap region with a width of ${\sim}10~\AU$
\citep{2010A&A...511A..75B,2011A&A...531A...1T,2013A&A...557A..68M}.
The inner dust likely consists of dust grains with size around one micron
($[0.1,~5]~\mum$ and $n=-3.5$ in \citealt{2010A&A...511A..75B} and \citealt{2011A&A...531A...1T},
$[0.1,~1.5]~\mum$ in \citealt{2013A&A...557A..68M}
).
A narrower gap region has been found in HD~139614 \citep{2014A&A...561A..26M,2016A&A...586A..11M}
from ${\sim}2.5~\AU$
to ${\sim}5.7~\AU$.
The dust composition of the inner dust in their {\RT} model is
$20\%$ graphite with small grain size ($[0.05,~0.2]~\mum$)
and
$80\%$ silicate with large grain size ($[5,~20]~\mum$).
Due to the much larger opacity of the small grains,
the optical-NIR opacity is dominated by the small graphite grain
and the silicate is introduced purely for interpreting the N-band silicate feature.
For HD~144432 \citep{2016A&A...586A..54C},
a even narrower gap between ${\sim}0.3~\AU$ and ${\sim}1.4~\AU$ was found,
and the NIR emitting material is likely optically thin dust consisting of mainly sub-micron grains.

Comparing our model A with the results for the three objects listed above,
there might be a trend that
objects with wider gaps tend to
have larger grain size in their inner disks.
However, a decisive conclusion cannot be made with this small sample.
Moreover, we cannot strictly exclude the possibility that
the inner dust in HD~169142 is optically thick,
with small grain size, and containing refractory species.
To confirm whether there is a trend for grain size to grow simultaneously with disk gap size,
more objects are needed to be similarly studied.

\fi

\section{Summary and conclusions}
\iftrue

In this paper we present a study of the dust properties
in the innermost sub-$\AU$ region in the disk of the Herbig star HD~169142,
using interferometric observations of the object with VLTI/PIONIER.
The object has been found to be variable in the NIR,
and has evolved from a pre-2000 state with higher NIR excess
to a post-2000 state with lower NIR excess.
The VLTI/PIONIER observations were performed in the 2011 and 2013,
when the object was already in low state.
We performed {\GM}ing and {\MCRT} modeling of the data,
complemented with SED data from literature.
The following results were obtained.

Both our {\GM}ing and the {\MCRT} modeling
indicate that the NIR-emitting hot dust is located at
${\sim}0.07~\AU$ from the central star.
Our {\MCRT} modeling shows that the SED and
visibilities can be simultaneously reproduced
with an optically thin dust model with large grain size (${\ge}1.0~\mum$),
or with an optically thick component at similar location.
Optically thin dust containing smaller grains
is excluded by our modeling,
because such a dust at the location of ${\sim}0.07~\AU$
would be overheated to a higher temperature (compared to gray dust),
therefore causing the modeled SED to be bluer than the observed one.
If the dust is optically thick,
then it must contain refractive species,
so that the hot surface layer is not destroyed.
Grain size distribution is not constrained in the optically thick case.

Based on the modeling,
we discussed the possible scenarios for the object
to evolve from its pre-2000 state (with higher NIR excess)
to its current state.
In our scenario 1, the inner dust is optically thin
in the pre-2000 state
and lost part of its mass later.
In scenario 2, the inner dust consists of an optically thin component and an optically thick core in the pre-2000 state,
and then lost the optically thin component later.
In both scenarios, the NIR variability arises solely from
changes affecting the optically thin component.
We argue that further IR photometric and interferometric observations will help to reduce the ambiguities in the current study.

Statistical studies of dependence of dust grain size distribution
on other disk parameters
can provide helpful information
for understanding
disk evolution.
By comparing our results with other interferometric studies
that constrained the dust grain size in the inner region of pre-transitional disks,
there seems to be a positive correlation between gap width
and grain size of inner dust.
However, more observations on larger samples are needed
for a firm conclusion.

\fi

\iftrue
\begin{acknowledgements}
This work was supported by the Momentum grant of the MTA CSFK Lend\"ulet Disk Research Group.
The authors are thankful to the generous help and useful comments from Myriam Benisty.
The comments from an anonymous referee helped to improve the quality of the paper.
\end{acknowledgements}
\fi

\bibliography{bib/instrument,bib/YSO,bib/other,bib/StarModel,bib/RADMC3D,bib/dust}

\iftrue
\appendix
\section{Detailed description of the fitting process with {\MCRT} models }

\subsection{Fitting with model type A (optically thin inner disk)}

\SetToScanAA
\StandardTableAndFigure
In our model scanning run \RunNameShort,
we tested {\ModelNumber} models with wide ranges of inner disk parameters.
We focused on testing different values of $L_*$, $R_\mathrm{in}$,
$\tau$, $n$, and $f_\mathrm{carbon}$, 
and kept other parameters fixed.
The parameter ranges and results 
are shown in \StandardTableAndFigureList.
The model {\ModelNameShort} approximately reproduces the whole data set
with $\chi^2_\mathrm{red}=\BestChiSquare$.

\SetToScanAAF
We then tested models in narrower parameter ranges to further constrain the parameters.
Because we found that the model results are insensitive to $f_\mathrm{carbon}$,
from then on we fixed it to $f_\mathrm{carbon}=0.5$.
In our model scanning run \RunNameShort,
we tested {\ModelNumber} models.
The parameter ranges and results 
are shown in \StandardTableAndFigureList.
The model {\ModelNameShort} approximately reproduces the whole data set
with $\chi^2_\mathrm{red}=\BestChiSquare$.

\subsection{Parameter study on the effects of ridial width in model type A}
\SetToScanAW
\StandardTableAndFigure
The purpose of our model scanning run {\RunNameShort}
is to test the effects of altering the radial width $W_\mathrm{ring}$
of the inner optically thin dust.
In this run we fixed most of the parameters to those in model AF1,
and scan different values
only for $W_\mathrm{ring}$, $R_\mathrm{in}$, and $n$,
testing {\ModelNumber} models.
The parameter ranges and results 
are shown in \StandardTableAndFigureList.
A degenaracy between $W_\mathrm{ring}$ and $R_\mathrm{in}$
is seen in the results.
For any given $W_\mathrm{ring}$,
a value of $R_\mathrm{in}$ can be found so that
the distance of NIR-emitting dust to the central star,
on an average sense, is similar to that in model AF1.
All such ($W_\mathrm{ring}$, $R_\mathrm{in}$) pairs
will lead to similar model predictions, including
visibilities and overall color of the dust.
This ambiguity is related to the fact that
the available interferometric observations
only partially resolved the inner dust of HD~169142.
Despite this ambiguity,
the constraint on dust grain size is robust
and
insensitive to the choice of $W_\mathrm{ring}$
(see \StandardTableAndFigureList{}b).

\subsection{Fitting with model type B (optically thick inner disk)}

\SetToScanBA
\StandardTableAndFigure
In our model scanning run \RunNameShort,
we tested {\ModelNumber} models with wide ranges of inner disk parameters.
We focused on testing different values of $L_*$, $R_\mathrm{in}$,
$h_\mathrm{in}$, and $n$,
and kept other parameters fixed.
The parameter ranges and results 
are shown in \StandardTableAndFigureList.
The model {\ModelNameShort} approximately reproduces the whole data set
with $\chi^2_\mathrm{red}=\BestChiSquare$.

\SetToScanBAF
We then tested models in narrower parameter ranges to further constrain the parameters.
Because we found that the model results are insensitive to $n$ for this type of model,
we now fix it to $n=-3.5$.
In our model scanning run \RunNameShort,
we tested {\ModelNumber} models.
The parameter ranges and results 
are shown in \StandardTableAndFigureList.
The model {\ModelNameShort} approximately reproduces the whole data set
with $\chi^2_\mathrm{red}=\BestChiSquare$.

\fi

\end{document}